%% file: ms.tex
\documentclass[twocolumn]{aastex631}

\def\numorbit{16}
\def\fivesiglimmag{30}

\def\threesiglimmagopt{31.1}

\input{institutions.tex}

\begin{document}

\title{Flashlights: More than A Dozen High-Significance Microlensing Events of Extremely Magnified Stars in Galaxies at Redshifts $z=0.7$--$1.5$}

\author[0000-0002-0786-7307]{Patrick L. Kelly}
\email{plkelly@umn.edu}
\affiliation{\UMN}
\author[0000-0003-1060-0723]{Wenlei Chen}
\affiliation{\UMN}
\author{Amruth Alfred}
\affiliation{\HongKong}
\author[0000-0002-8785-8979]{Thomas J. Broadhurst}
\affiliation{\IKERBASQUE}
\author[0000-0001-9065-3926]{Jose M. Diego}
\affiliation{\Cantabria}
\author{Najmeh Emami}
\affiliation{\UMN}
\author[0000-0003-3460-0103]{Alexei V. Filippenko}
\affiliation{\Berkeley}
\author{Allison Keen}
\affiliation{\UMN}
\author[0000-0002-4490-7304]{Sung Kei Li}
\affiliation{\HongKong}
\author{Jeremy Lim}
\affiliation{\HongKong}
\author[0000-0002-7876-4321]{Ashish K. Meena}
\affiliation{\BenGurion}
\author[0000-0003-3484-399X]{Masamune Oguri}
\affiliation{\Chiba; \ChibaPhys}
\author{Claudia Scarlata}
\affiliation{\UMN}
\author[0000-0002-8460-0390]{Tommaso Treu}
\affiliation{\UCLA}
\author{Hayley Williams}
\affiliation{\UMN}
\author[0000-0002-6039-8706]{Liliya L.R. Williams}
\affiliation{\UMN}
\author{Rui Zhou}
\affiliation{\UMN}
\author[0000-0002-0350-4488]{Adi Zitrin}
\affiliation{\BenGurion}
\author{Ryan J. Foley}
\affiliation{\UCSC}
\author[0000-0001-8738-6011]{Saurabh W. Jha}
\affiliation{\Rutgers}
\author{Nick Kaiser}
\affiliation{\ENS}
\author{Vihang Mehta}
\affiliation{\UMN}
\author{Steven Rieck}
\affiliation{\UMN}
\author{Laura Salo}
\affiliation{\UMN}
\author[0000-0001-5510-2424]{Nathan Smith}
\affiliation{\Arizona}
\author{Daniel R. Weisz}
\affiliation{\Berkeley}



\begin{abstract} 
Once only accessible in nearby galaxies, we can now study individual stars across much of the observable universe aided by galaxy-cluster gravitational lenses. When a star, compact object, or multiple such objects in the foreground galaxy-cluster lens become aligned, they can magnify a background individual star, and the timescale of a magnification peak can limit its size to tens of AU.
The number and frequency of microlensing events therefore opens a window into the population of stars and compact objects, as well as high-redshift stars.
To assemble the first statistical sample of stars in order to constrain the initial mass function (IMF) of massive stars at redshift $z=0.7$--1.5, the abundance of primordial black holes in galaxy-cluster dark matter, and the IMF of the stars making up the intracluster light, we are carrying out a 192-orbit program with the {\it Hubble Space Telescope} called ``Flashlights,'' which is now only two-thirds complete owing to scheduling challenges.  We use the ultrawide F200LP and F350LP long-pass WFC3 UVIS filters and conduct two 16-orbit visits separated by one year. Having an identical roll angle during both visits, while difficult to schedule, yields extremely clean subtraction. Here we report the discovery of more than a dozen bright microlensing events, including multiple examples in the famous ``Dragon Arc'' discovered in the 1980s, as well as the ``Spocks'' and ``Warhol'' arcs that have hosted already known supergiants. The ultradeep observer-frame ultraviolet-through-optical imaging is sensitive to hot stars, which will complement deep {\it James Webb Space Telescope} infrared imaging.  We are also acquiring Large Binocular Telescope LUCI and Keck-I MOSFIRE near-infrared spectra of the highly magnified arcs to constrain their recent star-formation histories.
\end{abstract}



\section{Introduction} 
\label{sec:intro}


Extreme gravitational lensing magnification (up to several  thousand) of individual high-redshift stars by a foreground galaxy cluster, a newly discovered phenomenon \citep{kellydiegorodney18,rodneybalestrabradac18,chenkellydiego19,kaurovdaivenumadhav19,welchcoediego22,diegopascalekavanagh22,chenkellytreu22,diegomeenaadams22}, has the ability to address three major outstanding questions.
(a) What is dark matter? Despite decades of searches, the constituents of dark matter remain unidentified.
(b) How was the universe reionized? The magnifying power of galaxy-cluster lenses holds great potential
for studying faint, high-redshift galaxies, provided one can construct sufficiently accurate lens models (e.g., \citealt{bouwensoeschillingworth17}).
(c) How do stellar populations at redshift $z=1$--2 differ from those in the nearby universe? We have only been able to study the composite spectral energy distributions (SEDs) of their luminous stars.

To address the major questions outlined above, we are carrying out ultradeep {\it Hubble Space Telescope (HST)} observations of the six Hubble Frontier Field (HFF; \citealt{lotzkoekemoercoe17}) clusters. These targets -- Abell 2744, MACSJ0416.1-2403, MACSJ0717.5+3745, MACSJ1149.5+2223, Abell S1063, and Abell 370 -- are among the most powerful gravitational lensing clusters and have multiband optical through infrared (IR) {\it HST} imaging observations as part of the 840-orbit HFF program.
The HFF survey obtained optical imaging of each galaxy-cluster field in two-orbit visits spread across a period of approximately a month. For a single two-orbit visit using the ACS WFC F6060W filter, the 5$\sigma$ limiting AB \citep{okegunn83} magnitude was 28.1 within a $0\farcs2$-radius aperture  (28.7 for an ``optimal'' extraction), according to the Exposure Time Calculator (ETC)\footnote{\url{https://etc.stsci.edu/etc/input/acs/imaging/}}.  Considering a single week of HFF visits amounting to a total of six orbits, the limiting magnitude in the F606W filter, for example, was 28.7 (or 29.3 for an optimal measurement). 

\begin{figure*}
\centering
\includegraphics[angle=0,width=5.5in]{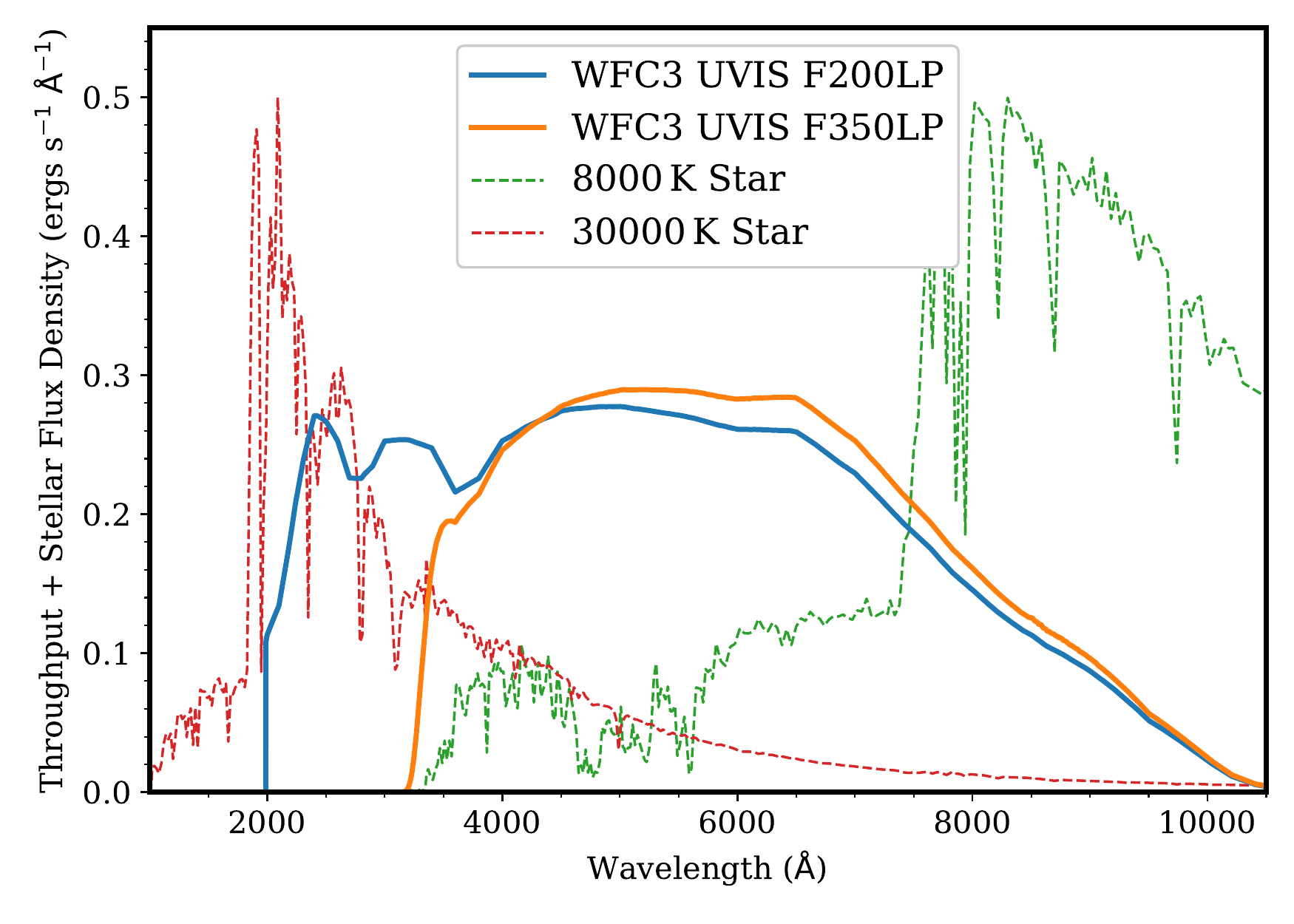}
\caption{The survey will observe each of six HFF galaxy clusters at two epochs. During each \numorbit-orbit visit, 
imaging alternates between orbits with integrations in WFC3 F200LP and WFC3 F350LP. The full throughputs of ACS WFC CLEAR and WFC3 F200LP
as well as the SEDs of two stars at $z=1$ are plotted here. The differences between the two filters provides constraints on the SED (including the reddening) of each star. } 
\label{fig:stars_filts}
\end{figure*}

Since the expected timescale for a $\sim 10$ R$_{\odot}$ star to cross a caustic can be only hours  \citep{miraldaescude91}, we aimed to carry out a two-epoch {\it HST} program of WFC3 UVIS  
observations (see Fig.~\ref{fig:stars_filts}) of  galaxy clusters with a single-visit 5$\sigma$ limiting magnitude of \fivesiglimmag\ (or optimal \threesiglimmagopt), $\sim 2$~mag deeper than each HFF visit, and $\sim 1.3$~mag deeper than each week of HFF visits. While the HFF obtained imaging from the optical through the near-infrared, our program includes observations with the ultraviolet-through-optical F200LP filter for sensitivity to emission from hot stars. 
The aim of the program is to find the first statistical samples of both microlensing events and pairs of magnified stellar images.  

Close to a so-called fold caustic, the area in the source plane with magnification exceeding a value $\mu$ scales as $\mu^{-2}$. By carrying out a survey with single-visit sensitivity $\sim 1.3$~mag (a factor of $\sim 3$ in flux) greater than that of a single week of HFF imaging, we require magnifications that are an equal factor of 3 smaller for detection. Consequently, to first order, a factor of $\sim 9$ greater number of stars through their microlensing events can be expected. For microlensing events with durations of only hours, the $\sim 2$\,mag difference in sensitivity corresponds to a factor of $\sim 36$.
The extremely deep imaging also enables
several additional high-impact investigations, and is needed to interpret {\it James Webb Space Telescope (JWST)}  observations of the HFFs.

In Section \ref{sec:scidrivers}, we describe the principal science motivation for the Flashlights survey. Four of the six target HFF galaxy-cluster fields have now been visited twice, and Section \ref{sec:microlensingevents} presents the high-significance microlensing events that we have detected. We discuss conclusions and implications from the current analysis of the data in Section \ref{sec:conclusions}. 



\section{Science Drivers}
\label{sec:scidrivers}

\subsection{Dark Matter}
Despite decades of searches, the constituents of $\sim 85$\% of the matter in the universe remain unidentified.
The newly discovered extreme magnification of high-redshift stars provides an entirely new and highly sensitive
probe of the nature of dark matter. First, the potential of a galaxy cluster acts to exaggerate the Einstein radii of objects in its intracluster medium by factors of up to $\sim 100$ near its critical curves, dramatically enlarging  microlensing ability \citep{diegokaiserbroadhurst18,venumadhavdaimiraldaescude17}. 
Second, light traveling through a cluster (or a galaxy) near its critical curve traverses an extreme density of dark matter. Consequently, microlensing fluctuations of background stars should reveal 
even a small fraction (1--2\%) of dark matter in the form of compact objects \citep{oguridiegokaiser18}. 

\begin{figure*}[th!]
\centering
\includegraphics[angle=0,width=7in, height = 3in]{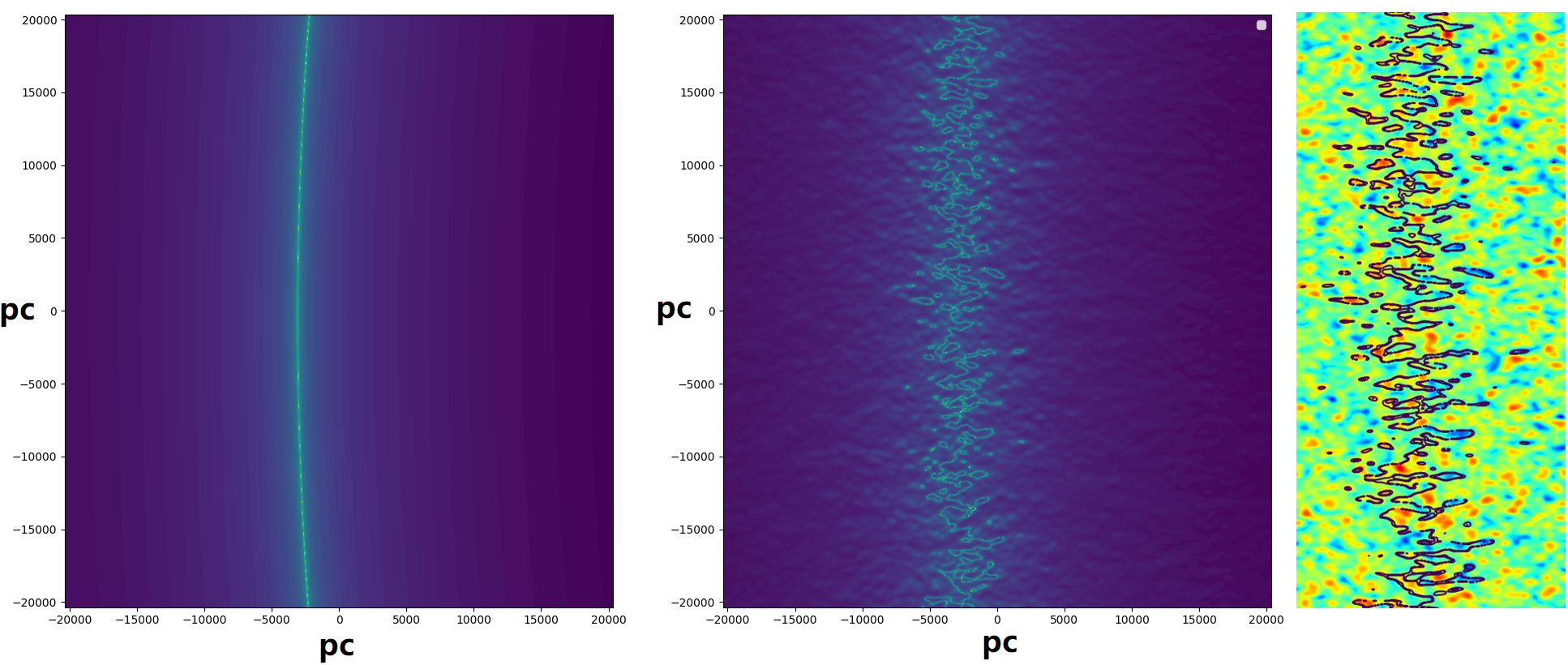}
\caption{Effect of wave dark matter on gravitational lensing magnification adjacent to the galaxy cluster's critical curve. {\it Left panel:} the magnification surrounding the critical curve of a galaxy cluster with mass similar to that of Abell 370 given a simply-parameterized smooth dark matter halo at the cluster scale and for individual cluster members. {\it Middle panel:} the magnification when the dark matter consists of ultralight bosons ($\psi$DM). {\it Right panel:} projected density fluctuations arising from $\psi$DM along with contours indicating regions of magnification of over $10^3$.}
\label{fig:darkmatter}
\end{figure*}


\begin{figure*}[th!]
\centering
\includegraphics[angle=0,width=4.0in]{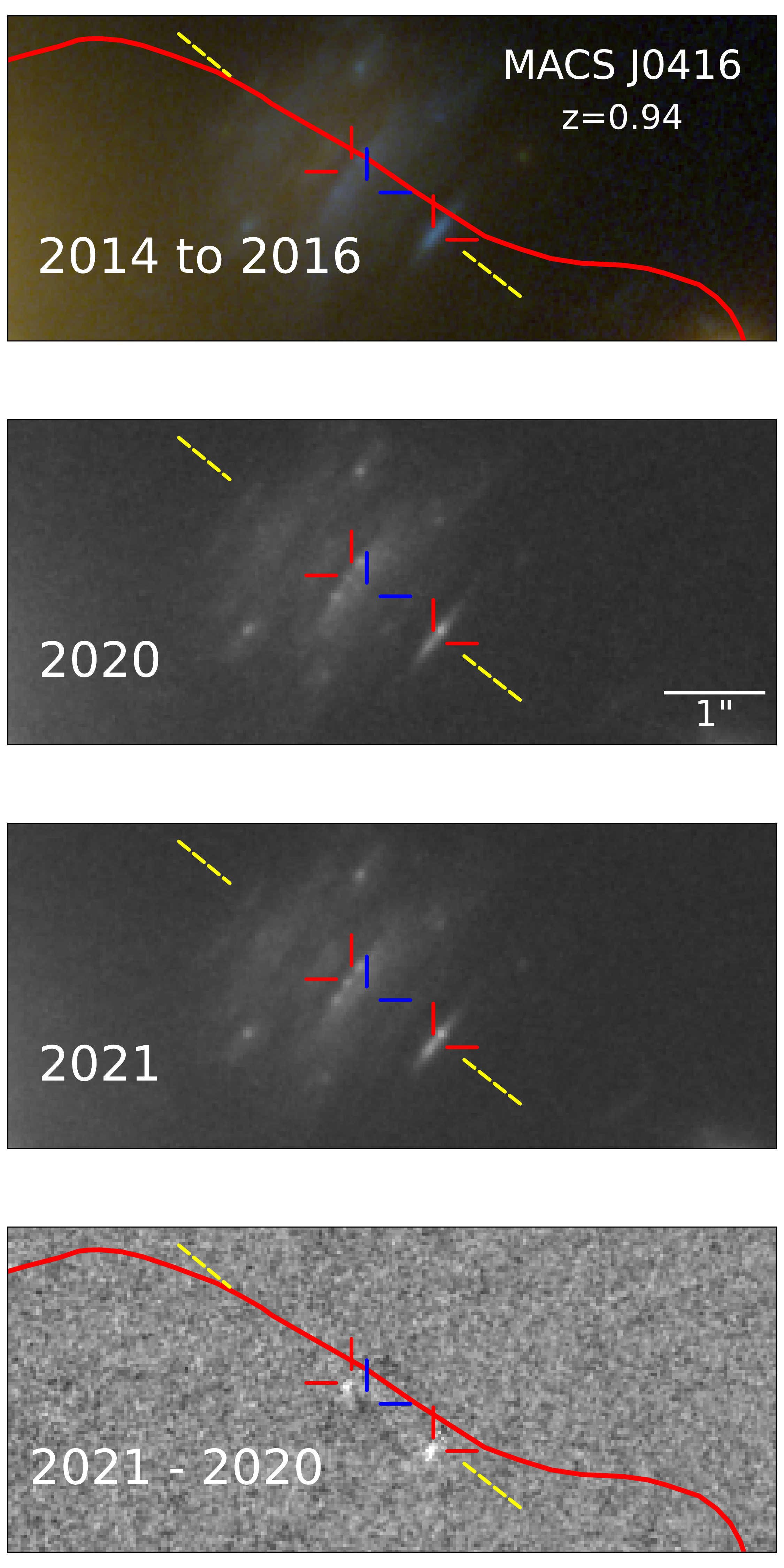}
\caption{Microlensing events within $< 0\farcs1$ of the galaxy cluster critical curve in an arc at $z=0.94$ in the MACS\,J0416 field. The upper panel shows a color-composite image from HFF imaging, the middle two panels show Flashlights F200LP imaging from 2020 and 2021, and the bottom panel indicates the difference of the two epochs.
The peak visible in the bottom panel that is farthest to the right corresponds to one of the images of a highly magnified blue supergiant, ``Warhol'' \citep{chenkellydiego19,kaurovdaivenumadhav19}.
Red hatches correspond to peaks detected in the 2021 imaging, and blue hatches correspond to those present in the 2020 imaging. Yellow dashed lines mark approximately the line of symmetry of the arc, which corresponds closely to the critical curve.  The red solid line marks the location of the galaxy cluster's critical curve according to the simply-parameterized Keeton model (v4) available from the HFF website \citep{keeton10}, although the model does not use the pairs of knots visible in the arc as constraints.
\label{fig:m0416_warhol}}
\end{figure*}

\begin{figure*}[th!]
\centering
\includegraphics[angle=0,width=4.25in]{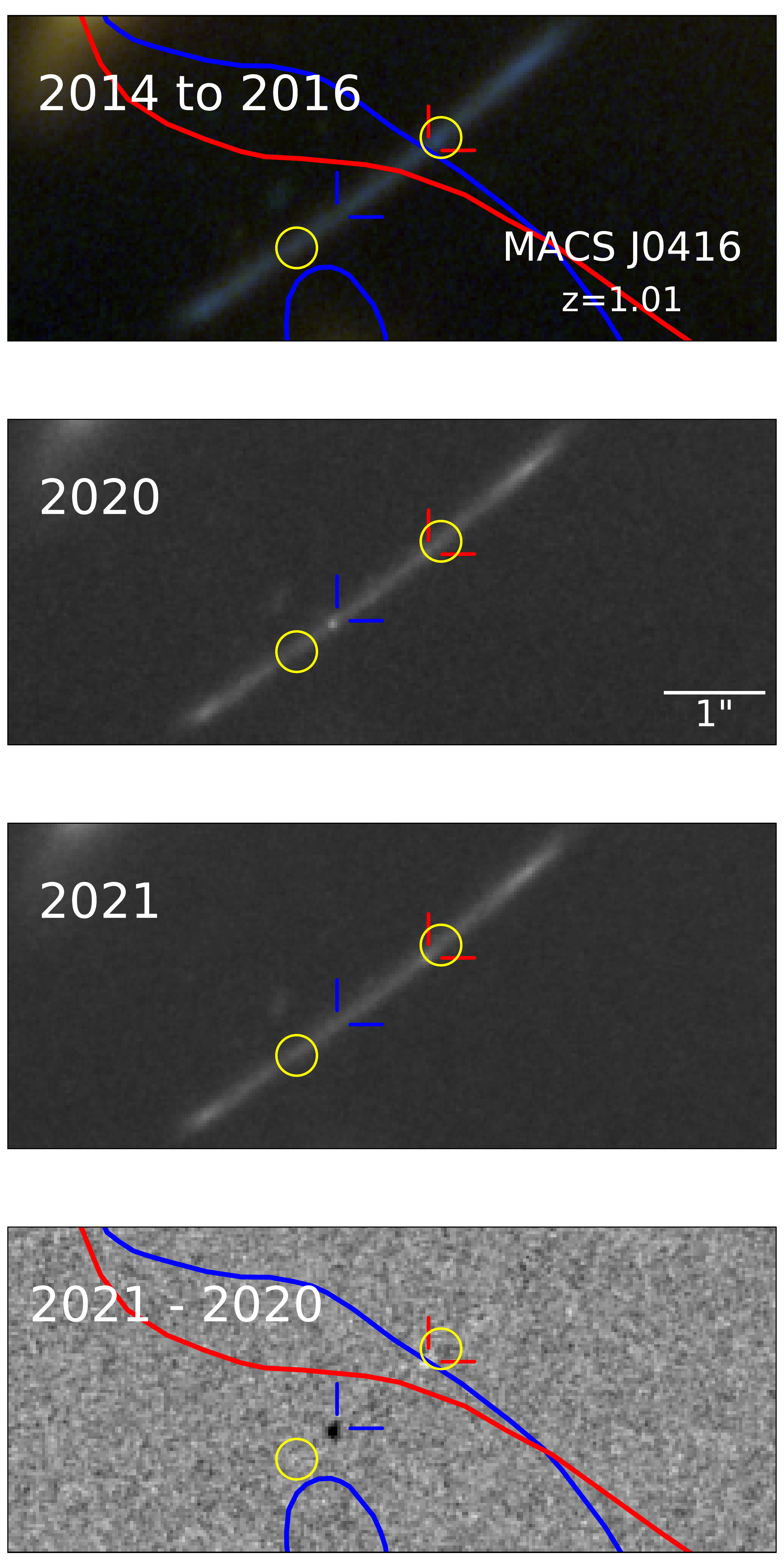}
\caption{Microlensing events in the ``Spocks'' arc at $z=1.00$ in the MACS\,J0416 galaxy-cluster field. We overplot the critical curve from the Zitrin Light-Traces-Mass model (v1) \citep{zitrinmeneghettiumetsu13} in red, and the Sharon \citep{johnsonsharonbayliss14} simply-parameterized model (v4cor) in blue. The locations of the two transients reported by \citet{rodneybalestrabradac18} in HFF imaging acquired in 2014 are shown with yellow circles. The positions of the two high-significance events detected in Flashlights imaging are at distinct positions, and may correspond to counterimages of the \citet{rodneybalestrabradac18} events.
\label{fig:m0416_spocks}}
\end{figure*}

\begin{figure*}[th!]
\centering
\includegraphics[angle=0,width=4.25in]{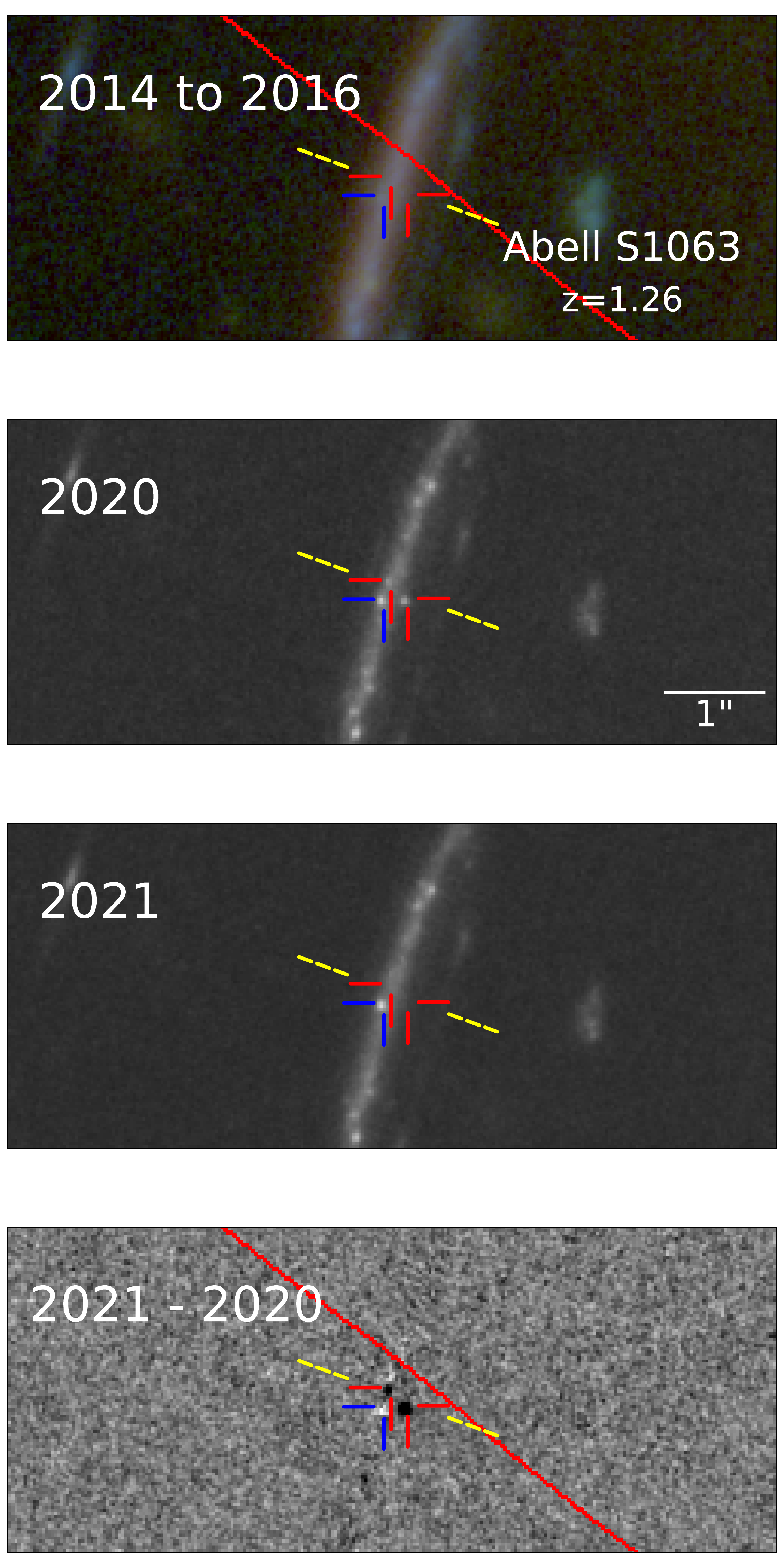}
\caption{Three high-significance microlensing events in a giant arc at $z=1.26$ in the Abell S1063 galaxy cluster field.  Yellow dashed lines mark approximately the line of symmetry of the arc, which corresponds closely to the critical curve. The plotted critical curve is that from the Keeton HFF simply-parameterized model (v4) \citep{keeton10}.
\label{fig:as1063_east}}
\end{figure*}

\begin{figure*}[th!]
\centering
\includegraphics[angle=0,width=4.25in]{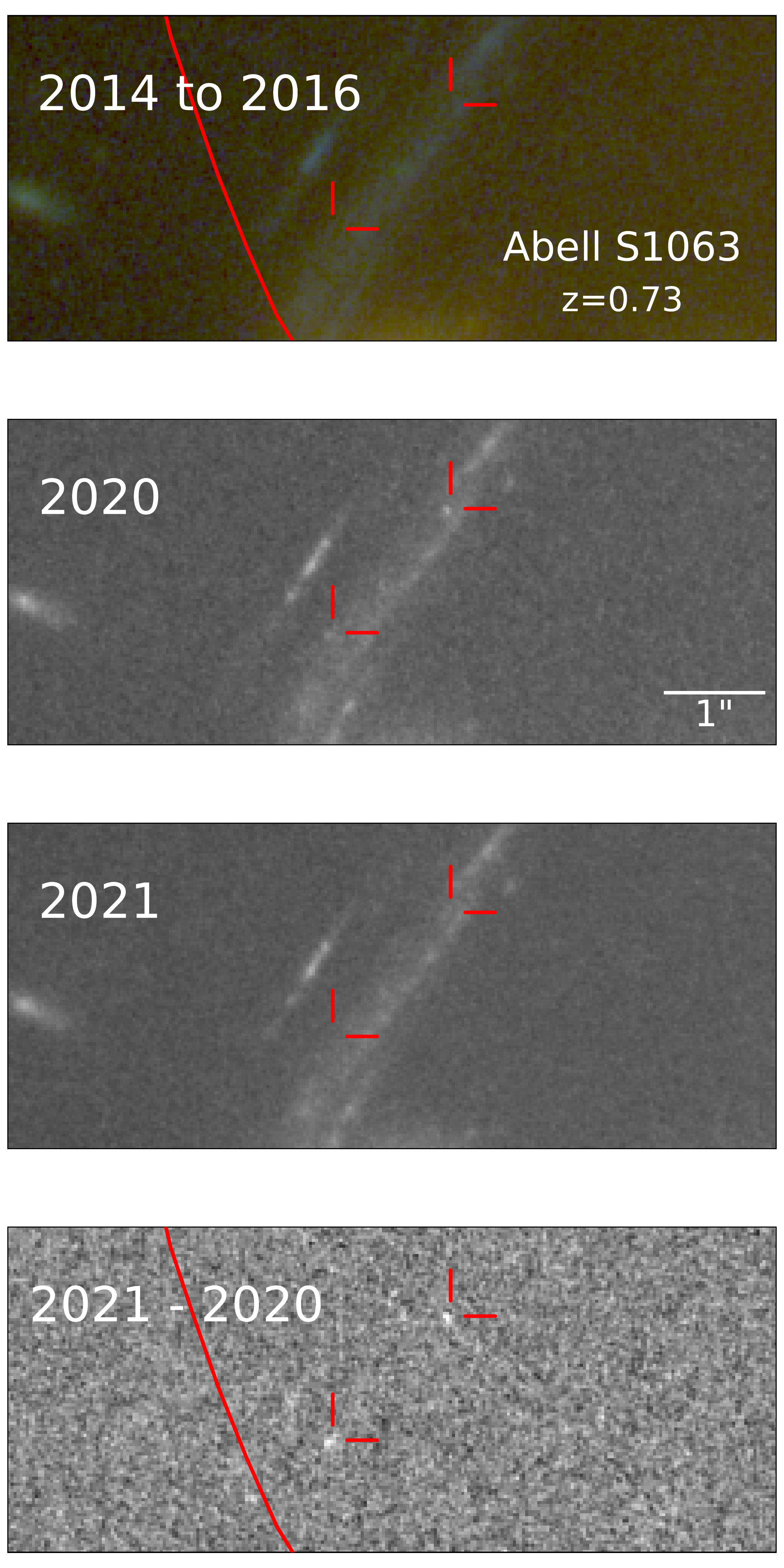}
\caption{Two high-significance candidate microlensing events in a giant arc at $z=0.73$ in the Abell S1063 galaxy cluster field.  The plotted critical curve is that from the Williams free-form {\tt Grale} \citep{ghoshwilliamsliesenborgs21} HFF model (v4.1).}
\label{fig:as1063_west}
\end{figure*}

\begin{figure*}[th!]
\centering
\includegraphics[angle=0,width=7.in]{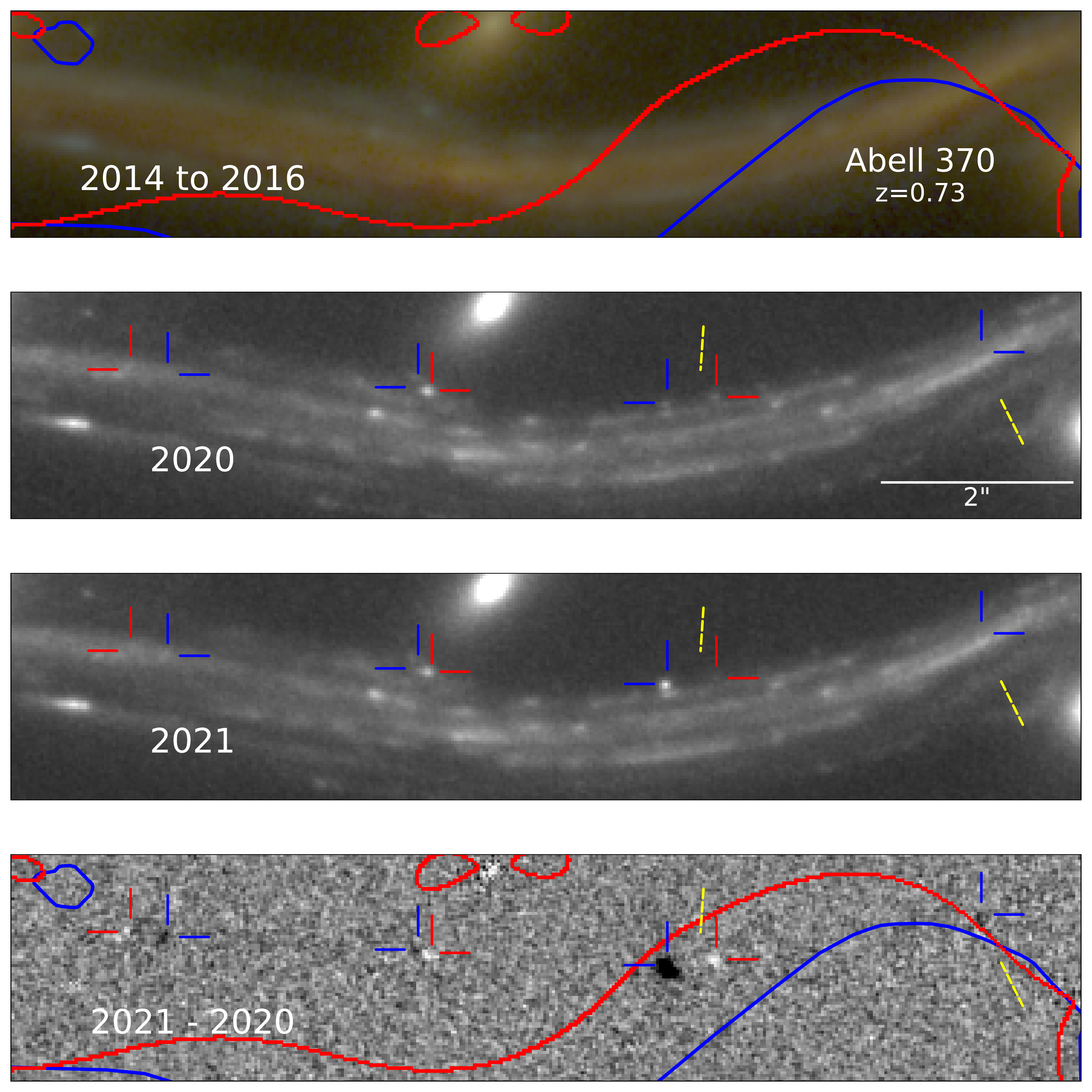}
\caption{High-significance microlensing events adjacent to a model critical curve in the  famous ``Dragon'' giant arc at $z=0.725$ \citep{soucailmellierfort88}.  The critical curve for the WSLAP+ model (v4.1) \citep{diegoprotopapassandvik05,diegotegmarkprotopapas07,diegobroadhurstwong16} is shown in blue, and that for the simply-parameterized {\tt GLAFIC} \citep{kawamataishigakishimasaku18,kawamataoguriishigaki16,oguri10} code (v4) is in red. The locations of the smaller critical curves associated with cluster members are not yet well constrained. 
\label{fig:a370}}
\end{figure*}


Primordial black holes (PBHs) could have formed from perturbations in (for example) an inflaton or spectator field in the early
universe. 
Currently,  Galactic \citep{tisserandleguillouafonso07,alcockallsmanalves01} and quasar \citep{mediavillajimenezvicentemunoz17} microlensing could allow PBHs with $10^{-3}$~M$_{\odot} < M < 10^2$~M$_{\odot}$ to account for up to 3--10\% of dark matter. Limits on their abundance place constraints on the primordial power spectrum associated with inflation (e.g., \citealt{carrtenkanenvaskonen17}).

The Laser Interferometer Gravitational-Wave Observatory (LIGO) is now able to detect the coalescence of neutron-star \citep{GW170817ligodiscovery17} and black-hole \citep{abbottabbottabbott16,LIGOGW151226} binary systems.  LIGO and surveys of X-ray binary systems, however, are not sensitive to the populations of either {\it single} black holes or binary systems with wide separations.  

At low optical depths for microlensing, the frequency of microlensing events should increase in approximately direct proportion to the mass density of microlenses. Moreover, the density of foreground intracluster stars along the line of sight toward giant arcs with $z=0.7$--1.5 that straddle the critical curve  are $\sim 1$\% of the projected density of dark matter.
Therefore, the number of microlensing events will provide a straightforward test of the abundance of PBHs comprising $\sim 1$\% of dark matter across an extremely broad mass range of at least $10^{-5} < m/{\rm M}_{\odot} < 100$. 


When a strongly lensed star lies close to a fold caustic in the source plane, its two most highly magnified images will appear as a {\it pair} straddling the critical curve.
When an object in the cluster becomes temporarily aligned with a background lensed star,
one of these images brightens (or fades) temporarily as has been observed for both Icarus \citep{kellydiegorodney18}
and Warhol \citep{chenkellydiego19,kaurovdaivenumadhav19}.
\citet{daivenumadhavkaurov18} show that the presence of subhalos such as those expected in 
the $\Lambda$CDM paradigm should imprint detectable $0\farcs03$--$0\farcs1$ deflections in the positions of pairs of lensed images of background stars, corresponding to halos of $\sim 10^7$--$10^9$~M$_\odot$.  
Low-mass halos have been claimed to be identified in optical observations of galaxy-scale gravitational lenses \citep{vegettikoopmansbolton10,vegettilagattutamckean12} as well as at submillimeter wavelengths in the well-resolved Einstein ring galaxy SDP.81 \citep{hezavehdalalmarrone16}. A complementary statistical approach has found evidence for subhalos from analysis of flux anomalies \citep{nierenbergtreuwright14,gilmanbirrernierenberg20}. In principle, several magnified stars along a giant arc may be sufficient to detect the presence of such subhalos depending on the degree of tidal stripping. 

In the context of ``wave'' dark matter ($\psi$DM), consisting of ultralight bosons, substructure is pervasive, caused by interference on the de~Broglie scale within galaxy and cluster halos \citep{schivechiuehbroadhurst14,huiostrikertremainewitten17,moczvogelsbergerrobles17}. The de~Broglie wavelength sets the scale of substructure fluctuations, ranging from several parsecs for clusters to $\sim 1$\,kpc for dwarf galaxies for a boson mass of $\sim$10$^{-22}$ eV \citep{pozobroadhurstdemartino21} as the de~Broglie wavelength scales inversely with momentum. The effect of this density modulation for lensing is predicted to produce highly corrugated critical curves \citep{chanschivewong20}. As shown in Fig.~\ref{fig:darkmatter}, we can approximate the projected density field expected for a massive galaxy cluster using the same procedure for galaxies as described by Alfred et al. (2022, submitted). Notice in particular how small ``islands'' of high magnification are expected to be significantly offset from the location of a galaxy cluster's critical curve, surrounding individual positive and negative density fluctuations that are depicted as red and blue (respectively). 
Consequently, individual stars may be highly magnified at projected positions that are relatively far from the Einstein radius, providing a testable prediction of $\psi$DM. 

\subsection{Improving Galaxy Clusters as Tools}
Near the critical curves of galaxy clusters, {\it JWST} should be sensitive to sources as faint as $\sim 35$~AB~mag, 
making it possible in principle to measure the properties of the low-luminosity galaxies thought to drive reionization. In areas of high magnification, it is crucial to obtain sufficient information to constrain the lens models well enough to determine accurate luminosities and volumes. Achieving this goal requires deep high-resolution images to identify dozens of multiply-imaged systems, as well as spectroscopic confirmations of large numbers of them.

In the image plane on the sky, magnification falls steeply as the inverse of the distance from the critical curve. 
In regions that currently lack sufficient constraints, however, lens models can disagree about the location of the critical curve by of order an arcsecond, which is a principal factor that explains the uncertain magnification of some high-redshift objects. As Icarus and Warhol demonstrate, each of the expected pairs of highly magnified stellar images (and those of star clusters) will {\it pinpoint} the location of the critical curve.
Furthermore, the probability of a microlensing event with a given peak 
magnification should drop approximately as the inverse of the offset from 
the critical curve to the fourth power \citep{kellydiegorodney18}.
Therefore, the microlensing events will also trace critical curves, albeit not as exactly. 

\subsection{Stellar Populations in Galaxies at $z=1$--2}
Galaxies near cosmic noon (at $z=1$--2) are more intensely star forming and exhibit more extreme nebular emission-line ratios than nearby galaxies \citep[e.g.,][]{steidelrudiestrom14,sandersshapleykriek15}. 
In nearby galaxies, we can resolve individual massive stars, and thereby constrain their luminosity functions and the upper end of the initial mass function (IMF). 
Currently, however, we are only able to study the composite SEDs of $z=1$--2 stellar populations, which limits our ability to constrain the upper end of the IMF. Improved constraints are necessary, because luminous stars are responsible for driving the evolution of galaxies through their ionizing flux, winds, and energy input from supernovae. Moreover, the IMF is a key parameter that affects interpretation of the observed properties of galaxy populations at high redshift \citep[e.g.,][]{narayanandave12}.

The number of detections of microlensing events should depend on the stellar luminosity function (see Extended Data Fig. 5 of \citealt{kellydiegorodney18}), and therefore the IMF.  
Consequently, the microlensing events detected by Flashlights will make it possible to measure the stellar luminosity function and upper end of the IMF across star-forming $z=1$--2 galaxies that cross the critical curve.

The survey includes two visits to each cluster consisting of \numorbit\ contiguous orbits (or as close as possible),
alternating between orbits with WFC3 UVIS F350LP and WFC3 UVIS F200LP integrations. Our initial strategy was to use ACS WFC CLEAR instead of WFC3 UVIS F350LP, but the point-spread function (PSF) was degraded in our first observations, of Abell 370.
The two visits to each HFF cluster are separated by one year, so that we can take 
parallel field observations at the same roll angle which
will serve as a control experiment.

A microlensing peak should have a duration $R/v$, where $R$ is the size of the lensed
source and $v$ is the transverse velocity of the lensing system \citep{miraldaescude91}. 
Given the $\sim 1000$\,km\,s$^{-1}$ expected transverse velocity of the cluster lens system,
the several-week timescale of Icarus' May 2016 microlensing peak \citep{kellydiegorodney18} implies that 
the source only extends for at most several tens of AU. 

\subsection{Enabling {\it JWST} Searches for Dropout Galaxies}
A key objective of {\it JWST} is to answer the question of how the universe became reionized at $z > 6$
(e.g., \citealt{castellanoamorinmerlin16}). One main observable required to study the history of reionization is the
rest-frame UV galaxy luminosity function, which constrains the total star-formation rate (SFR) at any given redshift.
The principal technique for measuring the luminosity functions of high-redshift galaxies is to identify Lyman-break ``dropout" galaxies. The ionizing continuum (below 912\,\AA) of galaxies is strongly absorbed by neutral H~I gas in the interstellar medium and in the surrounding circumgalactic and intergalactic medium. Additionally, the continuum between 912 and 1216~\AA\ is attenuated by the Ly$\alpha$ forest, created in discrete systems along the line of sight \citep{giavalisco02}.

Typical Lyman-break colors (i.e., color difference between wavelengths redder and bluer than Ly$\alpha$) used to select $z>6$ galaxies amount to $\gtrsim 1$~mag (e.g., \citealt{finkelstein16}, for a review). The Lyman break, however, can be mimicked by low-redshift faint galaxies showing intrinsically red colors, including from a strong Balmer break, across the filters used to perform the selection. These red low-redshift interlopers include $z \approx 2$ passive galaxies, dusty star-forming objects, or strong emission-line starbursting dwarfs  (e.g., \citealt{livermoretrentibradley18}, \citealt{ateksianascarlata11}). A classic example is the putative $z \approx 11$ object identified by \citet{laportepello11} as a {\it J}-band dropout.  In follow-up work, \citet{hayeslaportepello12} demonstrated spectroscopically that the object in fact has $z \approx 2.1$, either a young heavily reddened starburst or a maximally old system with a very pronounced 4000\,\AA\ break. Had AB $\approx 30$ optical imaging (blueward of the Lyman break) been available, this low-redshift contaminant could have been identified \citep{hayeslaportepello12}.

Reaching AB magnitudes of \fivesiglimmag, the observations provide six benchmark fields for high-redshift galaxy searches with {\it JWST}. Specifically, the data will be unique to identify the population of $z \approx 2$ galaxies with colors mimicking those of $z = 7$--12 candidates. 

\begin{table*}
\begin{center}
\begin{tabular}{cccccccccc} 
 \hline
Cluster & RA & Dec & $z$ & F200LP & F350LP & $\mu$ & $\mu$ & Approx. CC \\
 & (deg.)  & (deg.)  &  & Signif. & Signif. & {\tt GLAFIC}* & {\tt Gravlens}* &  Proximity \\
 \hline\hline
Abell 370 & 39.9704121 & -1.5848997 & 0.73$^{\rm a}$ & 19.0 &  ... & 1700 & 5800 & $\lesssim$0.1$''$ \\ 
Abell 370 & 39.9718583 & -1.5848133 & 0.73$^{\rm a}$ & 5.2 &  ... & 60  & 26 & $\lesssim$0.3$''$ \\ 
Abell 370 & 39.9711217 & -1.5848492 & 0.73$^{\rm a}$ & 3.3 &  ... &  30  & 28 & $\lesssim$0.3$''$ \\ 
Abell 370 & 39.9702708 & -1.5848842 & 0.73$^{\rm a}$ & 4.7 &  ... &  170 & 89 & $\lesssim$0.1$''$ \\ 
Abell 370 & 39.9719888 & -1.5848139 & 0.73$^{\rm a}$ & 4.5 &  ... &  50 & 15 & $\lesssim$0.3$''$ \\ 
Abell 370 & 39.9710843 & -1.5848642 & 0.73$^{\rm a}$ & 5.1 &  ... &  25  & 20 & $\lesssim$0.3$''$ \\ 
Abell 370 & 39.9691621 & -1.5846231 & 0.73$^{\rm a}$ & 5.3 &  ... &  15 & 10 & $\lesssim$0.1$''$ \\ 
Abell 370 & 39.9672042 & -1.5849367 & 1.26$^{\rm b}$ & 5.9 &  ... &  44 & 60 & $\sim$0.6$''$ \\ 
Abell S1063 & 342.1925167 & -44.5304881 & 1.26$^{\rm c}$ & 5.8 &  3.3 &  110  & 100 & $\lesssim$0.1$''$\\ 
Abell S1063 & 342.1924193 & -44.5304881 & 1.26$^{\rm c}$ & 31.0 &  16.3 & 190 & 180 & $\lesssim$0.3$''$\\ 
Abell S1063 & 342.1924847 & -44.5304337 & 1.26$^{\rm c}$ & 7.1 &  3.5 & 250 & 210 & $\lesssim$0.2$''$\\ 
Abell S1063 & 342.1894900 & -44.5290447 & 0.73$^{\rm d}$   & 7.2 &  4.9 & 28 & 32 & $\lesssim$1$''$\\ 
Abell S1063 & 342.1890363 & -44.5286928 & 0.73$^{\rm d}$ & 8.2 &  0.0 & 20 & 27 & $\lesssim$3$''$\\ 
Abell S1063 & 342.1952641 & -44.5279212 & 1.23$^{\rm e}$  & 10.4 &  8.0 & 550 & 450 & $\lesssim$0.1$''$\\ 
MACS\,J0416 & 64.0388904 & -24.0701557 & 1.01$^{\rm f}$ & 18.7 &  12.7 & 55 & 77 & $\lesssim$0.2$''$ \\ 
MACS\,J0416 & 64.0386110 & -24.0699715 & 1.01$^{\rm f}$  & 18.7 &  12.7 & 78 & 1200 & $\lesssim$0.2$''$ \\ 
MACS\,J0416 & 64.0363046 & -24.0675050 & 0.94$^{\rm g}$  & 10.2 &  8.3 & 25 & 60 & $\lesssim$0.1$''$\\ 
MACS\,J0416 & 64.0365524 & -24.0673339 & 0.94$^{\rm g}$ & 5.9 &  6.7 & 30 & 70 & $\lesssim$0.1$''$\\ 
MACS\,J0416 & 64.0364386 & -24.0674422 & 0.94$^{\rm g}$ & 2.7 &  4.5 & 21 & 50 & $\lesssim$0.1$''$\\ 
MACS\,J0416 & 64.0365133 & -24.0673656 & 0.94$^{\rm g}$ & 2.6 &  2.6 & 29 & 60 & $\lesssim$0.1$''$\\ 
 \hline
 \end{tabular}
\caption{Examples of high-significance transients in strongly lensing arcs. (a) \citet{soucailmellierfort88}; (b) \citet{lagattutarichardclement17}; (c) \citet{caminhagrillorosati16}; (d) \citet{diegobroadhurstwong16}; (e) \citet{karmancaputicaminha17}; (f) \citet{rodneybalestrabradac18}; (g) \citet{caminhagrillorosati17}. \label{tab:detections}
 }
\end{center}

\end{table*}

\section{High-Significance Microlensing Events}
\label{sec:microlensingevents}

In Table~\ref{tab:detections}, we list  high-significance detections of transients in a set of giant arcs in the Flashlights targets Abell 370, MACS\,J0416, and Abell S1063 with repeat observations acquired as part of the Flashlights survey. There, we describe whether the transients are found adjacent ($\lesssim 0\farcs2$) to the critical curve of the galaxy cluster, and therefore are very likely to be microlensing events in regions of high magnification exceeding $\sim 100$. A star adjacent to (and on the correct side of) a fold caustic in the source plane will appear as a pair of highly magnified images. The relative time delay between such a pair of highly magnified images is only of order days.  Given the typical timescales of outbursts of weeks to months of massive stars, any strong asymmetry between the fluxes of a pair of images can only be explained by stellar microlensing. Many of the microlensing events appear in pairs across the critical curve, and simulations by \citet{kellydiegorodney18} show that the same star is often responsible for microlensing events on both sides of the critical curve.

Fig.~\ref{fig:m0416_warhol} shows the  prominent arc at $z=0.94$ in the MACS\,J0416 galaxy-cluster field where the  ``Warhol'' \citep{chenkellydiego19,kaurovdaivenumadhav19} event was discovered. The clear fold symmetry of the Warhol arc at $z=0.94$ makes it possible to identify the location of the critical curve which corresponds to the line of symmetry, and we detect three microlensing events in the arc. 
The middle two panels show 2020 and 2021 F200LP Flashlights imaging, respectively, while the bottom panel shows a difference image made by subtracting the imaging acquired in 2021 from that taken in 2020. The red and blue cross hatches indicate the locations of the high-significance transients.
We overplot, as an example of a simply parameterized prediction, the critical curves for Light-Traces-Mass (v1) \citet{zitrinmeneghettiumetsu13} and the simply parameterized Sharon (v4cor) \citet{johnsonsharonbayliss14} models available on the HFF website\footnote{\url{https://archive.stsci.edu/prepds/frontier/lensmodels/}}. The microlensing events include detection of variability at location of the pair of images of the ``Warhol'' blue supergiant. 

In Fig.~\ref{fig:m0416_spocks}, we also plot our detection of two transients in the arc in the MACS\,J0416 field of the ``Spocks'' ($z=1.01$) \citep{rodneybalestrabradac18} events. The locations of the two new events identified by Flashlights differ from those of the Spocks events whose positions we identify.

Fig.~\ref{fig:as1063_east} shows high-significance microlensing events that we detect in Flashlights imaging close to the critical curve of a giant arc in the Abell\,S1063 galaxy-cluster field at $z=1.26$. In the case of this arc, the location of the galaxy cluster's critical curve is constrained by the mirrored positions of knots. Finally, we detect two high-significance transients in an adjacent arc in this field with a lower redshift of 0.73. Fig.~\ref{fig:as1063_west} plots the locations of these transients in F200LP imaging, and the critical curve we show is that from the Williams free-form {\tt GRALE} \citep{ghoshwilliamsliesenborgs21} HFF model (v4.1).

The upper panel of Fig.~\ref{fig:a370} shows a color-composite image of the ``Dragon'' arc at $z=0.73$ \citep{soucailmellierfort88} constructed from HFF optical and near-IR imaging taken in 2014--2016. Superimposed are the critical curves of the WSLAP+ model (v4.1) \citep{diegoprotopapassandvik05,diegotegmarkprotopapas07,diegobroadhurstwong16} and the {\tt GLAFIC} model (v4) \citep{kawamataishigakishimasaku18,kawamataoguriishigaki16,oguri10}.  Three of the transients in the Dragon arc are very likely to be microlensing events, given their close proximity to the critical curve of the galaxy cluster. The four additional transients, which appear in pairs, may also be microlensing events if they are located sufficiently close to the critical curves of nearby cluster members.  Finally, in \citet{meenachenzitrin22}, we present an analysis of a transient in an additional arc at $z=1.26$ in the Abell 370 galaxy-cluster field that we infer to likely be a microlensing event, despite a relatively large offset from the critical curve. 

\section{Conclusions}
\label{sec:conclusions}
We show that sufficiently deep exposures reveal many microlensing events in individual galaxies at $z=0.7$--1.5. 
We anticipate that deep, repeat visits with {\it JWST} will reveal populations of highly magnified stars. Given its powerful near-IR sensitivity, {\it JWST} will detect much cooler stars, including luminous red supergiants.  At the same time, once completed, Flashlights will provide a complementary dataset with sensitivity to the population of hot stars near the peak of cosmic star formation. 

Through detections of the first substantial sample of highly magnified stars, Flashlights will enable a novel probe of the nature of dark matter as well as the IMF of luminous stars at $z=0.7$--1.5. The positions of stars provide a new constraint on galaxy-cluster models through accurate measurements of the location of the cluster critical curve. Finally, the extremely deep UV through optical wideband imaging provides a powerful means to identify low-redshift interlopers in Lyman-break dropout samples.

\bigskip
\bigskip


We thank Patricia Royale at the Space Telescope Science Institute (STScI) for her terrific effort scheduling the program, and our instrument scientist Annalisa Calamida for her careful review of the program.
This research was supported by NASA/{\it HST} grants GO-15936 and GO-16278 from STScI, which is operated by the Association of Universities for Research in Astronomy, Inc., under NASA contract NAS5-26555, and also by grant JPL-1659411 to support ground-based follow-up observations with the Keck-I telescope. A.V.F. is grateful for additional financial support from the Christopher R. Redlich Fund and numerous individual donors. P.L.K. acknowledges support through NSF grant AST-1908823, and NASA Keck JPL grant 1659411.
This work was supported by JSPS KAKENHI grants JP22H01260, JP20H05856, and JP20H00181.
Work by S.K.L. and J.L. is supported by the Collaborative Research Fund under grant C6017-20G, which is issued by the Research Grants Council of Hong Kong S.A.R.
A.K.M. and A.Z. acknowledge support by grant 2020750 from the United States-Israel Binational Science Foundation (BSF) and grant 2109066 from the United States National Science Foundation (NSF), and by the Ministry of Science \& Technology, Israel.

Some of the data presented herein were obtained at the W. M. Keck
Observatory, which is operated as a scientific partnership among the
California Institute of Technology, the University of California, and
NASA; the observatory was made possible by the generous financial
support of the W. M. Keck Foundation. We acknowledge the HFF Lens Model website that is hosted by the Mikulski Archive for Space Telescopes (MAST) at STScI.

%

\vspace{5mm}
\facilities{HST (WFC3); Large Binocular Telescope (LUCI); Keck (MOSFIRE)}


\software{Drizzlepac \citep{gonzaga12}; Source Extractor \citep{bert96}}




\bibliography{ms}
\bibliographystyle{aasjournal}



\end{document}

%% file: institutions.tex
\newcommand{\Chiba}{Center for Frontier Science, Chiba University, 1-33 Yayoi-cho, Inage-ku, Chiba 263-8522, Japan}
\newcommand{\ChibaPhys}{Department of Physics, Graduate School of Science, Chiba University, 1-33 Yayoi-Cho, Inage-Ku, Chiba 263-8522, Japan}

\newcommand{\HongKong}{Department of Physics, The University of Hong Kong, Pokfulam Road, Hong Kong}
\newcommand{\ENS}{Laboratoire de Physique de l’Ecole Normale Supérieure, ENS, Université PSL, CNRS, Sorbonne Université, Université de Paris, Paris, France}

\newcommand{\UCLA}{Department of Physics and Astronomy, University of California, Los Angeles, CA 90095}

\newcommand{\Berkeley}{Department of Astronomy, University of California, Berkeley, CA 94720-3411, USA}

\newcommand{\Arizona}{Department of Astronomy, University of Arizona, Tucson, AZ 85721, USA}
\newcommand{\Rutgers}{Department of Physics and Astronomy, Rutgers, The State University of New Jersey, Piscataway, NJ 08854, USA}

\newcommand{\Cantabria}{IFCA, Instituto de F\'isica de Cantabria (UC-CSIC), Av. de Los Castros s/n, 39005 Santander, Spain}

\newcommand{\IKERBASQUE}{Ikerbasque Foundation, University of the Basque Country, DIPC Donostia, Spain}

\newcommand{\UCSC}{Department of Astronomy and Astrophysics, UCO/Lick Observatory, University of California, 1156 High Street, Santa Cruz, CA 95064, USA}

\newcommand{\UMN}{Minnesota Institute for Astrophysics, University of Minnesota, 116 Church Street SE, Minneapolis, MN 55455, USA}

\newcommand{\BenGurion}{Physics Department, Ben-Gurion University of the Negev, P.O. Box 653, Beer-Sheva 8410501, Israel}